# Hypothesis: Muon Radiation Dose and Marine Megafaunal Extinction at the end-Pliocene Supernova

Adrian L. Melott[a], Franciole Marinho[b], and Laura Paulucci[c]

a Department of Physics and Astronomy, University of Kansas, Lawrence, Kansas 66045 USA.  melott@ku.edu

b Universidade Federal de São Carlos, Rodovia Anhanguera, km 174, 13604-900, Araras, SP, Brazil. fmarinho@ufscar.br c Universidade Federal do ABC, Av. dos Estados, 5001, 09210-170, Santo André, SP, Brazil.  laura.paulucci@ufabc.edu.br

ABSTRACT

Considerable data and analysis support the detection of one or more supernovae at a distance of about 50 pc, ~2.6 million years ago. This is possibly related to the extinction event around that time and is a member of a series of explosions which formed the Local Bubble in the interstellar medium. We build on previous work, and propagate the muon flux from supernova-initiated cosmic rays from the surface to the depths of the ocean. We find that the radiation dose from the muons will exceed the total present surface dose from all sources at depths up to a kilometer and will persist for at least the lifetime of marine megafauna. It is reasonable to hypothesize that this increase in radiation load may have contributed to a newly documented marine megafaunal extinction at that time.

Key words: Muons.  Supernovae. Mass extinctions. Megafauna.

1. Introduction

There is now considerable confirmatory evidence from multiple sources for a supernova (SN) or series of up to ten SN thereof over the last ~8 My (Knie *et al*., 2004; Fry *et al*., 2016; Mamajek, 2016; Melott, 2016; Wallner *et al*., 2016; Breitschwerdt *et al.*, 2016; Binns *et al*., 2016; Fimiani *et al.*, 2016; Ludwig *et al*., 2016; Erlykin *et al*., 2017). The strongest signal is of an event about 2.5 Ma, which coincides within the uncertainties to the Pliocene-Pleistocene boundary at which there was elevated extinction (hereafter PP), qualifying as a mass extinction by at least one set of criteria (Melott and Bambach, 2014). This extinction included a major megafaunal extinction which was more severe in coastal waters (Pimiento *et al*., 2017).

The possible terrestrial effect of a nearby SN has long been of interest. A major modern attempt to computationally approach the problem was published by Gehrels et al. (2003), and emphasized the effect of stratospheric ozone depletion and concurrent UVB increase at the Earth's surface as a mechanism for change. Benitez et al. (2002) noticed the proximity of the SN timing to the PP extinction and proposed the UV mechanism as related to the extinction. However, there was also awareness of the effect of muons on the subsurface from cosmic ray showers (Marinho *et al*., 2014). All of these works were limited by a relative lack of information on source electromagnetic and cosmic ray (CR) emissions. More recently, new information on prompt and early emissions from SNe as well as the timing and distance of relatively recent proximate SNe were used as input in a series of computations (Melott *et al*,

2017). We use terrestrial muon flux from this study as an input in order to propagate effects into the ocean.

## 2. Base study methods and conclusions

The work here is a follow-on to a recent fairly comprehensive study (Melott *et al*, 2017) modeling the best available information on the event indicated by recent data (Knie *et al.*, 2004; Fry *et al.*, 2016; Mamajek, 2016; Melott, 2016; Wallner *et al.*, 2016; Breitschwerdt *et al.*, 2016; Binns *et al.*, 2016; Fimiani *et al.*, 2016; Ludwig *et al.*, 2016). We will summarize the assumptions, methods, and results here, but the reader is referred to the original publication (Melott *et al*, 2017) for detailed information. A SN of type IIP is assumed to have taken place about 2.6 Ma, at a distance of 50 pc. We assume that the SN injects instantaneously the CR energy $2.5 \times 10^{50}$ erg with source spectrum $Q(E) \sim E^{-2.2} \exp(-E/E_c)$ and cutoff energy $E_c = 1$ PeV. We further assume that both the SN and the Earth are within the Local Bubble, having been already formed by SNe earlier in the series (Breitschwerdt *et al.*, 2016). The magnetic field in the Local Bubble is assumed to have been largely expelled and residing within the walls of this structure; the internal remnant field is turbulent with strength $B = 0.1$ µG. Under these conditions the cosmic ray propagation is essentially diffusive and no major anisotropy in the direction of arrival is expected. A power law spectrum was associated with the turbulent magnetic field which leads to a diffusion coefficient of cosmic rays after leaving the supernova remnant to behave as $D(E) \propto E^{1/3}$ for $E \ll E_c$. All this results in a cosmic ray intensity of I ∝ $Q(E)/(Dt)^3 \exp(-(-r/Dt)^2)$. Our results imply a much larger cosmic ray flux than estimated in a previous study (Gehrels *et al.*, 2003) which lacked information recently available. The flux we find at 50 pc is comparable to, but has higher energy cosmic rays than assumed in that previous study. The effects are correspondingly greater at the Earth's surface and in the oceans.

A second step was the convolution of the GCR spectrum with an updated version of the tables from Atri & Melott (2011) designed for the calculation of the muon fluxes on earth's surface. Those tables were originally obtained by performing simulations of the propagation of primary particles with fixed energy on the top of the atmosphere using the CORSIKA software (Pierog *et al*) which is a specific code for simulating air showers which allows the use of the EPOS model for high energy hadronic interactions and FLUKA model for the low energy hadron-nucleons interactions ($\lesssim$100GeV). A conservative lower end of 10GeV was chosen for the proton primaries as simulation results are known to match available atmospheric muon flux data well (Hebbeker & Timmermans, 2002).

It is interesting to note that the Gauss-Matuyama boundary (Cox & Dalrymple, 1967), when a reversal of the magnetic poles on earth took place, also happened around 2.6 Ma ago meaning that the geomagnetic field at the time could be substantially lower than the current value in the first few thousand years (Clement, 2004) after the SN event. This would increase the amount of low energy muons (up to ~3GeV) at the surface during this period, accounting for a higher dose than the one calculated here in the ocean's shallow waters (first ~15m of water) making a stronger case for the proposed hypothesis but not affecting our numerical estimates as depth increases.

It was found, based on more recent observations that the immediate flux of gamma rays, X-rays, and UV photons from 50 pc is too small to be significant. It was found that the flux of blue light from the remnant is large enough to be detrimental to wildlife if in the night sky, but only for a timescale of order one month, too short to be visible in the fossil record.

Cosmic ray propagation in the turbulent field leads to a time-dependent flux at the Earth. We reproduce here as Figure 1 the Figure 1B in Melott *et al*. (2017). The units of this figure are adjusted so that equal areas under the curve correspond to equal energy flux at the Earth. The zero of time is taken to be the arrival of the first prompt photons at the Earth. Finally, the boundary conditions are taken to be open, so that the CR continue past the Earth. However, in the case of the Local Bubble, the CR should reflect off the boundary of the region, and establish an approximately steady-state inside the region. The flux shown exceeds the current "galactic cosmic ray" flux at the Earth by more than two orders of magnitude at PeV energies.

For many decades, the dominant effect considered for terrestrial life from a nearby SN was ozone depletion. High energy photons and CR can ionize and dissociate molecules in the atmosphere, notably $N_2$. Through a complicated chain of chemical reactions (Thomas, 2005), the resultants catalyze the depletion of $O_3$ with a recovery timescale of 5-10 years after the ionizing flux ends. In the recent study, the CR flux was taken to be constant at the level of 100 years. In this case, a globally averaged ozone depletion of order 25% was established as the equilibrium result (Melott *et al.*,

2017). It was estimated that this would result in an approximately 50% increase in UVB at the surface. This is much larger than the few percent increase induced by chlorofluorocarbons in the recent past. It would no doubt be detrimental, but not a mass extinction level event. Rainout would contribute a slow but steady addition of nitrate fertilizer to the ground (Gehrels *et al*., 2003). Ionization in the troposphere might contribute to an increase in cloud-to-ground lighting (e.g. Chilingarian *et al.,* 2017), but consideration of this is beyond the scope of this article.

    Radiation on the ground will consist primarily of neutrons and muons. Although the neutron flux increased, the amount at ground level would constitute a very small radiation dose. The tables of Atri and Melott (2011) were convolved with the resultant cosmic ray spectra to determine muon flux on the ground. This muon dose was found to increase by a factor of about 150 at the 100 year CR flux level, and would build up to about 1 Sv per 30 years, which would have a substantial effect on the cancer and mutation rates. Overall, this is ~14 times higher than the average total radiation dose from all sources currently experienced on earth's surface. A quantitative risk assessment for marine megafaunal species due to the CR flux level increase is not possible as there are no specific studies available. However, the risk estimation as a function of dose is a relatively developed procedure for human beings in cases of doses above several 100 mSv with acute and chronic exposures for solid tumors. For instance, the NCI radiation risk assessment tool indicates an excess lifetime risk of developing cancer of 16-40% with respect to the expected baseline future risk for a 1 Sv chronic exposure over 30 years with a 90% C.L. (Gonzalez et al., 2012). Larger and longer-lived organisms will experience a greater relative increase in radiation dose,

which may be related to a marine megafaunal extinction at the Pliocene-Pleistocene boundary where 36% of the genera were estimated to become extinct (Pimiento *et al*., 2017).

We note that muons will dominate the dose at the ground and the ocean (O'Brien *et al.*, 1998; Simonsen *et al*., 2000; UNSCEAR Report, 1996). They produce the largest ionizing dose of any secondary component from cosmic rays (Marinho *et al*., 2014) and this effect will increase with depth in the ocean. Therefore it is appropriate to focus on the muon dose.

## 3. Computing the muon dose in the oceans

In order to estimate the radiation dose caused by cosmic ray muons as a function of the depth in the oceans we employed a 3D Monte Carlo approach to characterize the energy deposition of these particles along their trajectories, using data from (Melott *et al.*, 2017) as the initial conditions at the surface of the oceans. For this we used the Geant4 framework (Agostinelli *et al.*, 2003) which provides all functionalities necessary to simulate the radiation-matter interactions in water with a few percent precision and over the whole energy range considered (Bogdanov *et al*., 2006). Muons can transfer part of their energy to the material directly or via production of secondary particles close to their trajectory through ionization, Bremsstrahlung, pair production and nuclear processes which were all taken into account in our calculations. In this method the trajectories can be resumed by the occurrence of the particle decay or due to capture at rest. As shown in (Marinho *et al*., 2014), the generated secondaries are mostly electrons and gammas and they will likely cause a lateral displacement of ~1m so still typically inside the large volumes we are interested in.

The geometry adopted for the simulation was a water column with 100km of height and 30km wide. The starting muon positions were homogeneously sampled within a horizontal area of 2 x 2 km² centered on top of the column to ensure particles always propagating within the material and to avoid unwanted boundary effects. This choice of geometry does not introduce any effect of significance considered its dimensions with respect to earth's curvature. Propagation directions were sampled according to the muon angular distributions on earth's surface (Patrignani *et al.*, 2017). After propagating the muons through the material and fixing all with the same initial energy E, the average deposited energy per particle ΔE'(E,z) was calculated in smaller volumes positioned at different z depths in the water column. These were block volumes of area A = 1 x 1 km² also centered in the column z-axis and with a thin h = 1m height such that the dose rate due to a flux ϕ(E) could be obtained as function of depth. The dose rate was then calculated as:

$$D(z) = \int_{E_{min}}^{E_{max}} \frac{\Delta E'(E,z)}{h\rho} \frac{d\phi}{dE} dE,$$

where $\rho$ is the water density and $E_{min}$ and $E_{max}$ configure the energy range of the fluxes as from (Melott *et al.*, 2017).

## 4. Results of muon computations

Figure 2 shows the values obtained for annual dose as function of depth. The solid black line without markers shows the typical annual dose at the present time. The dose from muons in the ocean exceeds the present dose for 10 kyr with open boundary conditions on the cosmic rays from the SN. It even exceeds the average dose from all sources at the Earth's surface for up to 1 kyr depending on depth—for depths from 100 m to 1 km. Again, with the CR confining boundary of the Local Bubble, the dose can be expected to last longer.

Muons' dose can also be characterized as a function of their energy at a certain depth instead of the initial energies at the surface. Figure 3 presents the differential dose as a function of energy for when the flux reaches its maximum (100 years after the supernova event) for two different oceanic depths and also at the surface. The differential dose due to the present muon flux at the surface is also shown for comparison. The small contribution to the dose from the very low energy part of the spectrum (<1GeV) increases with depth surpassing the surface contribution as can be observed at $z = 250$ m. This is due to the muons energy loss as they traverse water, shifting muons at mid energy range to lower energies. This was also noticed in a similar analysis in (Marinho *et al.*, 2014). However, this contribution will decrease faster with higher depths such that it goes below the surface curve again at $z = 500$ m. This energy loss also renders the flattening and decrease of both underwater curves observed up to ~100 GeV.

The muon stopping power is within the same order of magnitude across the whole energy range considered in this analysis (Melott *et al.*, 2017) with a steep increase only below the minimum and above the maximum energy values shown (Bogdanov *et al.*, 2006; Patrignani *et al.*, 2017). The simulation indicates that only muons arriving at water surface with energy above ~64 GeV and ~125 GeV reach $z = 250$ m and $z = 500$ m depths, respectively. This is in agreement with the stopping power estimate of ~2.5 MeV/cm for these energies. Muons with energies above 200 GeV contribute with 24%, 42%, and 52% of the deposited dose for z=0, 250m, and 500m, respectively, after 100 years from the SN event. Therefore, high energy muons can reach deeper in the oceans being the more relevant agent of biological damage as depth increases. In contrast, this contribution to the present dose at the surface amounts to only 7.7%.

## 5. Estimates of biological effects and the Pliocene-Pleistocene extinction

The PP event qualifies as a mass extinction by one quantitative analysis (Melott and Bambach, 2014), but analysis here is complicated by the fact that there were a number of simultaneous changes, including a marine megafaunal extinction (Pimiento *et al.*, 2017), extinction of bivalves in the western Atlantic area (Stanley, 1986), and extinctions, vegetation changes, and hominid evolution in Africa (deMenocal, 2004).

Causes generally associated with PP are climate change (many glaciations and a generally cooler climate in the Pleistocene) and the closing of the Isthmus of Panama as North and South America joined. The latter caused changes in ocean currents, affecting temperatures and nutrient levels in the Caribbean and western Atlantic Ocean. It also caused an exchange of fauna between North and South America. The two had evolved in isolation from one another, and many extinctions occurred as species out-competed one another. Deforestation in northeast Africa may have been related to climate change. For all these reasons, evaluation of effects from the SN is complicated. The possible rise on the ionization levels on the atmosphere due to the increased CR flux caused by the SN explosion may possibly be connected to climate change (Svensmark *et al.*, 2017), but this is a controversial claim. Here, we are concerned with the effects of the muons.

Muons normally constitute a variable approximate 10% of the radiation dose on the ground. Figure 2 shows that this dose goes up by more than two orders of magnitude as a result of secondaries from the SNCRs. The dose from various sources to organisms in the ocean should be less than on land, due to the short range of various emissions. Alphas from $^{22}$Rn, a primary source on land should be reduced. The mass fraction of this isotope in water above the thermocline (about the top km) is comparable to that in air (Broecker *et al.*, 1967). The muons, which are highly penetrating, will constitute a relatively much larger increase in radiation dose in the upper hundreds of meters of water than they will on land.

There will also be a differential effect based on the size of organisms. Again, most natural radiation is not penetrating, and will constitute a surface effect. Muon irradiation will effectively penetrate any organisms, and so constitute a volume effect. Larger organisms are largely self-shielded from most external natural radiation sources, but not from muons. The larger the organism, the larger will be the relative increase in radiation dose.

While it can be argued that the muons generally constitute a small increase in the radiation dose inducing mutations and cancer (Melott *et al.*, 2017), large organisms in the top few hundred meters of the ocean will experience a much larger relative increase in radiation dose during the period of SNCR irradiation. It is therefore tempting to associate this with the PP megafaunal extinction, which was concentrated in coastal waters (Pimiento *et al.*, 2017). Both large size and a habitat in the upper ocean would constitute high risk factors, as the organisms would experience a greater relative increase in radiation dose from the muons.

## Acknowledgments

Research support for ALM was provided by NASA under research grant NNX14AK22G. We thank the referees for their comments helpful in clarifying certain aspects of the manuscript.

## Disclosure

There are no competing financial interests.

## References

Agostinelli, S., Allison, J., Amako, K., Apostolakis, J., Araujo, H., Arce, P., Asai, M., Axen, D., Banerjee, S., Barrand, G., Behner, F., Bellagamba, L., Boudreau, J., Broglia, L., Brunengo, A., Burkhardt, H., Chauvie, S., Chuma, J., Chytracek, R., Cooperman, G., Cosmo, G., Degtyarenko, P., Dell'Acqua, A., Depaola, G., Dietrich, D., Enami, R., Feliciello, A., Ferguson, C., Fesefeldt, H., Folger, G., Foppiano, F., Forti, A., Garelli, S., Giani, S., Giannitrapani, R., Gibin, D., Gomez Cadenas, J.J., Gonzalez, I., Gracia Abril, G., Greeniaus, G., Greiner, W., Grichine, V., Grossheim, A., Guatelli, S., Gumplinger, P., Hamatsu, R., Hashimoto, K., Hasui, H., Heikkinen, A., Howard, A., Ivanchenko, V., Johnson, A., Jones, F.W., Kallenbach, J., Kanay, N., Kawabata, M., Kawabata, Y., Kawaguti, M., Kelner, S., Kent, P., Kimura, A., Kodama, T., Kokoulin, R., Kossov, M., Kurashige, H., Lamanna, E., Lampen, T., Lara, V., Lefebure, V., Lei, F., Liendl, M.,

Lockman, W., Longo, F., Magni, S., Maire, M., Medernach, E., Minamimoto, K., Mora de Freitas, P., Morita, Y., Murakami, K., Nagamatu, M., Nartallo, R., Nieminen, P., Nishimura, T., Ohtsubo, K., Okamura, M., O'Neale, S., Oohata, Y., Paech, K., Perl, J., Pfeiffer, A., Pia, M.G., Ranjard, F., Rybin, A., Sadilov, S., Di Salvo, E., Santin, G., Sasaki, T., Savvas, N., Sawada, Y., Scherer, S., Sei, S., Sirotenko, V., Smith, D., Starkov, N., Stoecker, H., Sulkimo, J., Takahata, M., Tanaka, S., Tcherniaev, E., Safai Tehrani, E., Tropeano, M., Truscott, P., Uno, H., Urban, L., Urban, P., Verderi, M., Walkden, A., Wander, W., Weber, H., Wellisch, J.P., Wenaus, T., Williams, D.C., Wright, D., Yamada, T., Yoshida, H., and Zschiesche, D. (2003) Geant4 - a simulation toolkit. *Nuclear Instruments and Methods in Physics Research A* 506(3):250–303.

Atri, D., and Melott, A.L. (2011) Modeling high-energy cosmic ray induced terrestrial muon flux: A lookup table. *Radiation Physics and Chemistry* 80:701-703.

Benitez, N., Maiz-Apellaniz, J., and Canelles, M. (2002) Evidence for Nearby SN Explosions. *Physical Review Letters* 88:081101.

Binns, W.R., Israel, M.H., Christian, E.R., Cummings, A.C., de Nolfo, G.A., Lave1, K.A., Leske, R.A., Mewaldt, R.A., Stone, E.C., von Rosenvinge, T.T., Wiedenbeck, M.E. (2016) Observation of the $^{60}$Fe nucleosynthesis-clock isotope in galactic cosmic rays. *Science* 352:677-680.

Bogdanov, A.G., Burkhardt, H., Ivanchenko, V.N., Kelner, S.R., Kokoulin, R.P., Maire, M., Rybin, A.M., and Urban, L. (2006) Geant4 simulation of production and interaction of Muons. *IEEE Transactions on Nuclear Science* 53(2):513-519.

Breitschwerdt, D., Feige, J., Schulreich, M.M., de Avillez, M.A., Dettbarn, C., and Fuchs, B. (2016) The locations of recent SNe near the Sun from modelling $^{60}$Fe transport. *Nature* 532:73-76.

Broecker, W.S., Li, Y.H., and Cromwell, J. (1967) Radium-226 and Radon 220: Concentration in Atlantic and Pacific Oceans. *Science* 158:1307-1310.

Chilingarian, A., Chilingaryan, S., Karapetyan, T., Kozliner, L., Khanikyants, Y., Hovsepyan, G., Pokhsrayan, D., and Soghomonyan, S. (2017) On the initiation of lightning in thunderclouds. *Scientific Reports* 7:1371. DOI:10.1038/s41598-017-01288-0.

Clement, B. M. (2004) Dependence of the duration of geomagnetic polarity reversals on site latitude. *Nature* 428: 637-640.

Cox, A., and Dalrymple, G.B. (1967) Geomagnetic polarity epochs; Nunivak Island, Alaska:Earth and Planetary Science Letters 3:173-177.

deMenocal, P. B. (2004) African climate change and faunal evolution during the Pliocene–Pleistocene. *Earth and Planetary Science Letters* 220:3-24.

Erlykin, A.D., Machavariani, S.K., and Wolfendale,A.W. (2017) The Local Bubble in the interstellar medium and the origin of the low energy cosmic rays. *Advances in Space Research* 59:748-750.

Fimiani, L., Cook, D.L., Faestermann, T., Gómez-Guzmán, J.M., Hain, K., Herzog, G., Knie, K., Korschinek, G., Ludwig, P., Park, J., Reedy, R.C, and Rugel G. (2016) Interstellar $^{60}$Fe on the Surface of the Moon. *Physcal Review Letters* 116:151104.

Fry, B.J., Fields, B.D., and Ellis, J.R. (2016) Radioactive iron rain: transporting $^{60}$Fe in SN dust to the ocean floor. *The Astrophysical Journal* 827:48.

Gehrels, N., Laird, C.M., Jackman, C.H., Cannizzo, J.K., Mattson, B.J., and Chen, W. (2003) Ozone Depletion from Nearby SNe. *The Astrophysical Journal* 585:1169-1176.

Gonzalez, A.B., Apostoaei, A.I., Veiga, L.H.S., Rajaraman, P., Thomas, B.A., Hoffman, F.O., Gilbert, E., and Land, C. (2012) RadRAT: a radiation risk assessment tool for lifetime cancer risk projection. *Journal of Radiological Protection* 32 (3):205.

Hebbeker, T., and Timmermans, C. (2002) A compilation of high energy atmospheric muon data at sea level, Astroparticle Physics 18 (1):107-127.

Knie, K., Korschinek, G., Faestermann, T., Dorfi, E.A., Rugel, G., Wallner, A. (2004) $^{60}$Fe Anomaly in a Deep-Sea Manganese Crust and Implications for a Nearby SN Source. *Physical Review Letters* 93(17):171103.

Ludwig, P., Bishop, S., Egli, R., Chernenko, V., Deneva, B., Faestermann, T., Famulok, N., Fimiani, L., Gómez-Guzmán, J.M., Hain, K., Korschinek, G., Hanzlik, M., Merchel, S., and Rugel G. (2016) Time-resolved 2-million-year-old SN activity discovered in Earth's microfossil record. In *Proceedings of the National Academy of Sciences* (PNAS) 113 edited by E. Bard, National Academy of Sciences, pp. 201601040.

Mamajek, E. E. (2016) A Pre-Gaia Census of Nearby Stellar Groups. In *Proc. IAU Symp. 314, Young Stars & Planets Near the Sun*, 21 edited by J. H. Kastner, B. Stelzer, and S. A. Metchev, Cambridge University Press, Cambridge, pp. 21-26.

Marinho, F., Paulucci, L., and Galante, D. (2014) Propagation and energy deposition of cosmic rays' muons on terrestrial environments. *International Journal of Astrobiology* 13:319-323.

Melott, A.L. and Bambach, R.K. (2014) Analysis of periodicity of extinction using the 2012 geological timescale. *Paleobiology* 40:177-196.

Melott, A.L. (2016) SNe in the Neighbourhood. *Nature* 532:40-41.

Melott, A.L., Thomas, B.C., Kachelrieß, M., Semikoz, D.V., and Overholt, A.C. (2017) A SN at 50 pc: Effects on the Earth's Atmosphere and Biota. *The Astrophysical Journal* 840:105. https://doi.org/10.3847/1538-4357/aa6c57

O'Brien, K., Friedberg, W., Smart, D.F., and Sauer H.H. (1998) The atmospheric cosmic - and solar energetic particle radiation environment at aircraft altitudes. *Advances in Space Research* 21(12):1739–1748.

Patrignani, C. *et al.* (Particle Data Group) (2017) Review of Particle Physics. *Chinese Physics C* 40:100001.

Pierog, T., Heck, D. and Knapp, J., CORSIKA (Cosmic Ray Simulations for Kascade) an air shower simulation. Program homepage: https://www.ikp.kit.edu/corsika/

Pimiento, C., Griffin, J.N., Clements, C.F., Silvestro, D., Varela, S., Uhen M.D., and Jaramillo, C. (2017) The Pliocene marine megafauna extinction and its impact on functional diversity. *Nature Ecology & Evolution* 1:1100-1106.

Simonsen, L. C., Wilson, J.W., Kim, M.H., and Cucinotta, F.A. (2000) Radiation exposure for human Mars exploration. *Health Physics: The Radiation Safety Journal* 79(5):515-525.

Stanley, S.M. (1986) Anatomy of a Regional Mass Extinction: Plio-Pleistocene Decimation of the Western Atlantic Bivalve Fauna. *Palaios* 1:17-36.

Svensmark, H., Enghoff, M.B., Shaviv N.J., and Svensmark. J. (2017) Increased ionization supports growth of aerosols into cloud condensation nuclei. *Nature Communications* 8:2199.

Thomas, B.C. (2005) Gamma-Ray Bursts and the Earth: Exploration of Atmospheric, Biological, Climatic, and Biogeochemical Effects. *The Astrophysical Journal* 634:509-533.

United Nations Scientific Committee on the Effects of Atomic Radiation (1966), Radiation from natural sources. In UNSCEAR report, Annex A, U. N., New York, pp. 6-49.

Wallner, A., Feige, J., Kinoshita, N., Paul, M., Fifield, L.K., Golser, R., Honda, M., Linnemann, U., Matsuzaki, H., Merchel, S., Rugel, G., Tims, S.G., Steier, P., Yamagata, T., and Winkler, S.R. 2016: Recent near-Earth SNe probed by global deposition of interstellar radioactive $^{60}$Fe. *Nature* 532:69-72.

**Figure Captions**

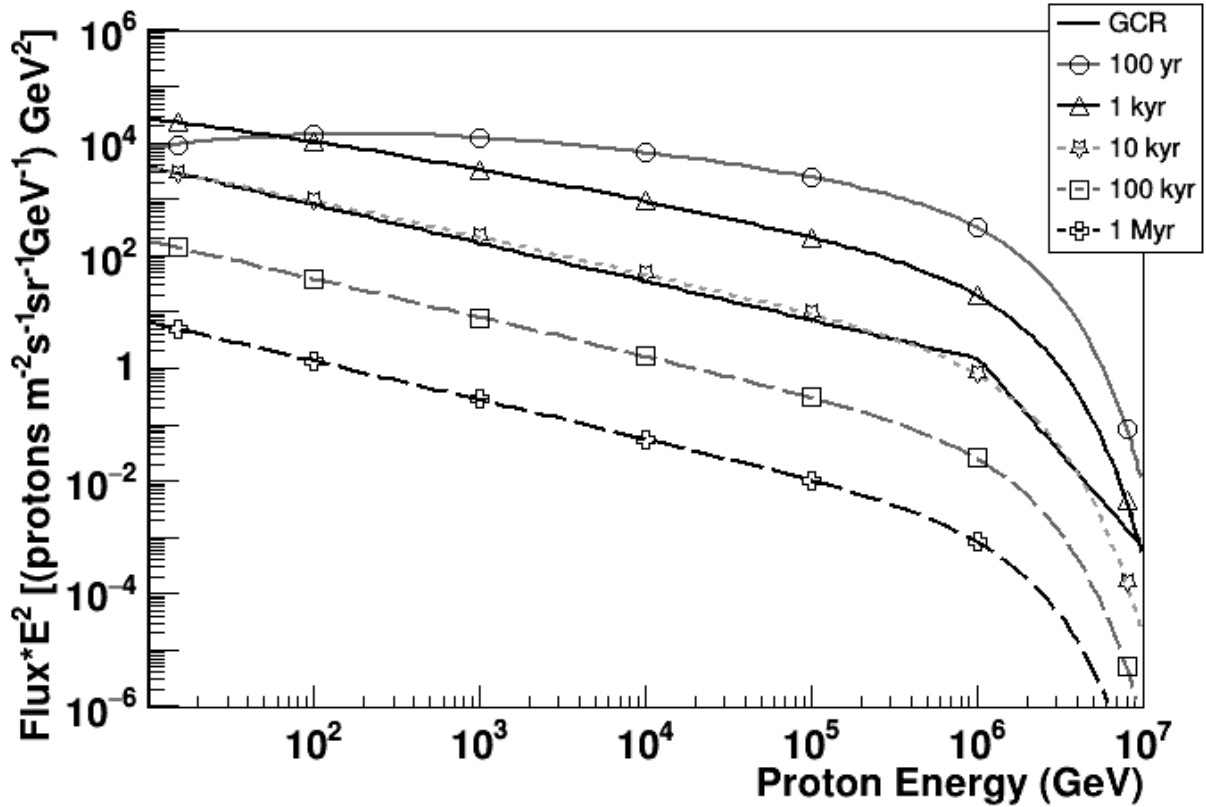

Figure 1. This is a very slightly modified reproduction of Figure 1B from reference Melott *et al.* 2017, the time-dependent cosmic ray flux at the Earth from a SN at 50 pc, propagating diffusively through a tangle 0.1 µG magnetic field. It takes 10,000 years for the flux to return to normal levels (solid black line). However, a steady-state would be reached inside the Local Bubble with magnetic field compressed into the walls. Reproduced courtesy American Astronomical Society.

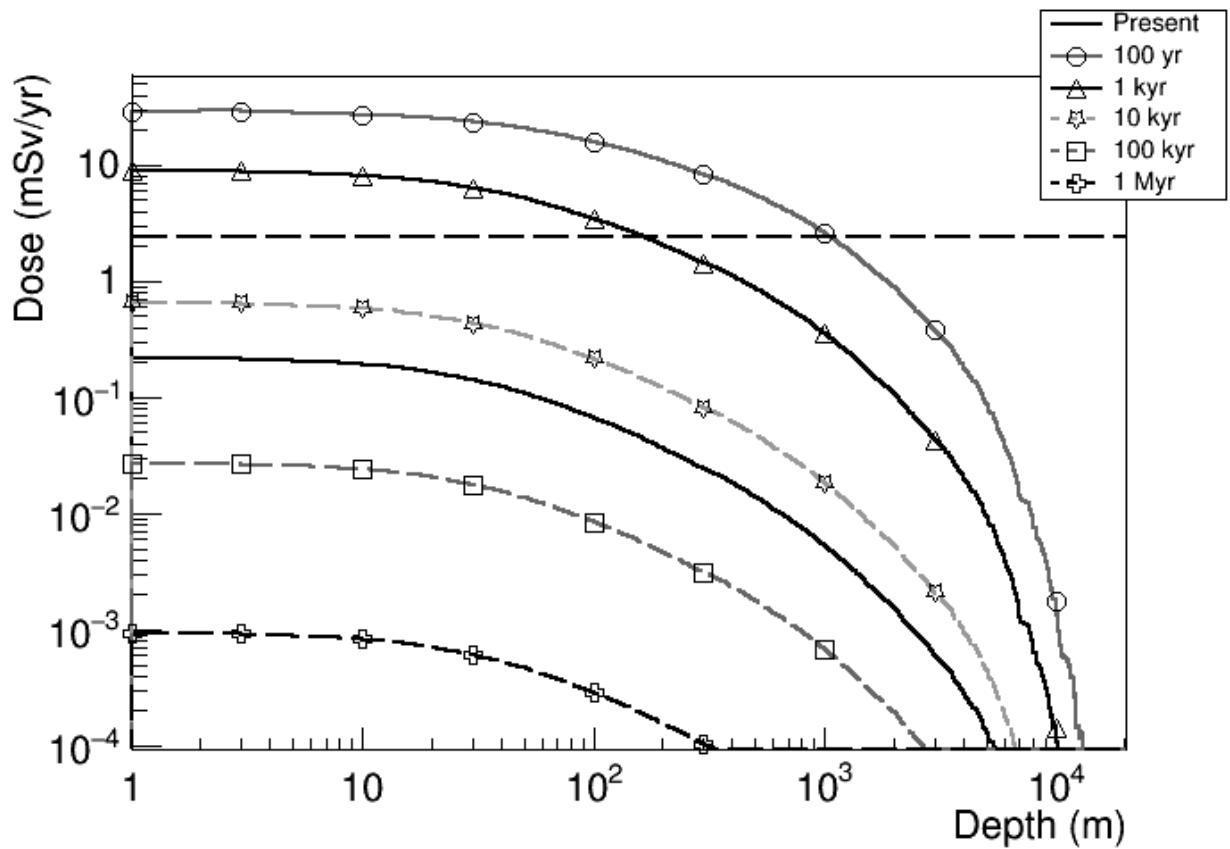

Figure 2: Estimated annual dose caused by muons in the oceans due to cosmic ray flux from a SN at 50 pc. The horizontal dashed line indicates the current total dose from all sources at Earth's surface while the solid black line without markers indicates current muon's dose in the ocean from the SN as function of depth.

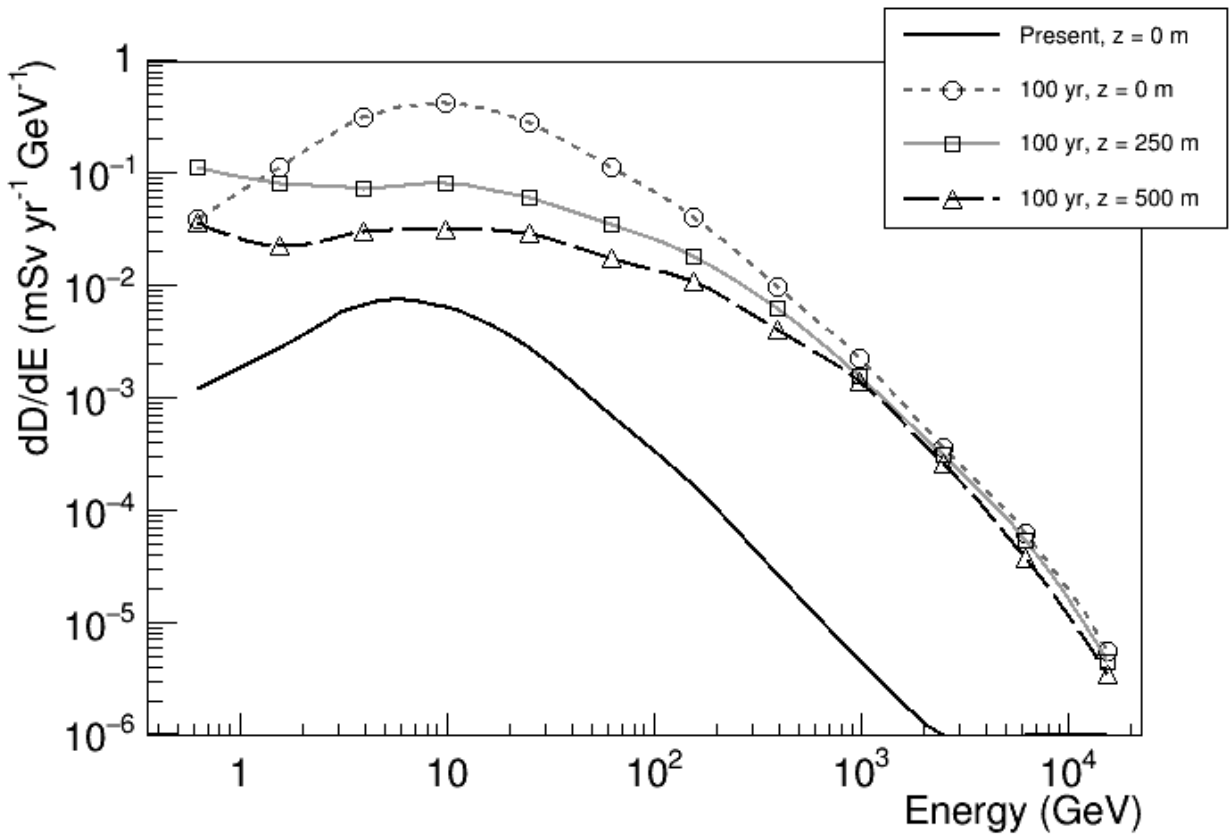

Figure 3: Estimated differential dose caused by muons in terrestrial oceans due to cosmic ray flux after 100 years from a SN at 50 pc distance for different depths. The present differential dose is shown for comparison.